# 3D Simulation of Superconducting Magnetic Shields and Lenses using the Fast Fourier Transform

**Leonid Prigozhin[1] and Vladimir Sokolovsky[2]**

[1]J. Blaustein Institutes for Desert Research, Ben-Gurion University of the Negev, Sede Boqer Campus 84990, Israel
[2]Physics Department, Ben-Gurion University of the Negev, Beer-Sheva 84105, Israel

E-mail: leonid@math.bgu.ac.il, sokolovv@bgu.ac.il

**Abstract.** Shielding sensitive scientific and medical devices from the magnetic field environment is one of the promising applications of superconductors. Magnetic field concentration by superconducting magnetic lenses is the opposite phenomenon based, however, on the same properties of superconductors: their ideal conductivity and ability to expel the magnetic field. Full-dimensional numerical simulations are necessary for designing magnetic lenses and for estimating the quality of magnetic shielding under arbitrary varying external fields. Using the recently proposed Fast Fourier Transform based three-dimensional numerical method [Prigozhin and Sokolovsky, ArXiv 1801.04869] we model performance of two such devices made of a bulk type-II superconductor: a magnetic shield and a magnetic lens. The method is efficient and can be easier to implement than the alternative approaches based on the finite element methods.

## 1. Introduction

Superconductor and hybrid superconductor/ferromagnet systems are able to efficiently shield stronger magnetic fields than the purely ferromagnet systems and to provide large magnetic field attenuations; see [1-4] and the references therein. Hollow superconductor cylinders, either open or closed at one end, are often employed to shield the area inside their hole. The efficiency of shielding has usually been estimated for a uniform external field either parallel to the cylinder axis [1-3] or perpendicular to it [1]. Since in the latter case the cylinder was assumed infinitely long, both configurations were described by two-dimensional models. The problem of shielding a non-uniform axisymmetric field could also be reduced to a two-dimensional one [5]. The only solution of a three-dimensional (3D) shielding problem we know was obtained in [4] for a general non-uniform field using the finite element software GetDP.

For a finite cylinder the problem is 3D also if the applied field is uniform but not parallel to the cylinder axis. Here we solve such a 3D problem for a hollow superconducting cylinder sealed at one end (the configuration considered in [2]) using the Fast Fourier Transform (FFT) based method [6]. This method is an easily implemented alternative to the finite element methods for 3D magnetization problems (see, e.g., [7-12]) and we employ it to model also field concentration by a magnetic lens. The configuration of a bulk superconductor in the latter problem is that of the magnetic lenses in [13]. Our work demonstrates applicability of the 3D FFT-based method to realistic magnetization problems arising in practical applications of bulk type-II superconductors.

Although here we assumed a field-independent isotropic power law current-voltage relation characterizing the superconductor material, our 3D method allows one to use a different relation to account for a critical current density dependence on the magnetic field, anisotropy of a superconductor, or the flux cutting effect in force free configurations. However, the method is currently not applicable to problems with transport current and/or infinitely long cables.

All numerical simulations in this work have been performed in Matlab R2017a on a PC with Intel(R) Core(TM) i5-2400 CPU @ 3.10GHz, 16 GB RAM, Windows 7 (64-bit). As an example, our program for the magnetic lens simulation can be downloaded from [14]. Adapting this program to a different geometry and another current-voltage relation should not be difficult.

## 2. FFT-based method for bulk magnetization problems

An FFT-based numerical method for solving superconductivity problems was first proposed for thin film magnetization problems in [15] and then applied to modeling thermal instabilities and flux avalanches in superconducting films; see, e.g., [16-19]. In our work [6] this 2D method was compared to the finite element methods and a new 3D FFT-based method for bulk magnetization problems was derived. As in the 2D thin film case, the 3D method employs FFT for the approximation in space and the method of lines for integration in time. However, the proposed method uses the magnetic field as the main variable and is based on a different problem formulation as follows.

Let $\Omega \subset R^3$ be the domain occupied by a superconductor, $\Gamma$ - its boundary, and $\boldsymbol{j}$ - the current density satisfying $\nabla \cdot \boldsymbol{j} = 0$ in $\Omega$ and having a zero normal component, $j_{\mathrm{n}} = 0$, on $\Gamma$. For simplicity, we assume the applied external magnetic field $\boldsymbol{h}_{\mathrm{e}}(t)$ is uniform and $\boldsymbol{j} = \boldsymbol{0}$ in $\Omega_{\mathrm{out}} = R^3 \setminus \bar{\Omega}$. We will restrict our consideration to the contourwise simply connected domains (every closed contour in the domain can be presented as a boundary of a surface also belonging to this domain) and refer to [6] for a discussion of the more complicated multiply connected case. As is most often done, we assume the electric field $\boldsymbol{e}$ and the current density $\boldsymbol{j}$ in the superconductor are parallel and

$$\boldsymbol{e} = \rho(|\boldsymbol{j}|)\boldsymbol{j}, \tag{1}$$

where $\rho(|\boldsymbol{j}|) = (e_0 / j_{\mathrm{c}})|\boldsymbol{j} / j_{\mathrm{c}}|^{n-1}$ is the nonlinear resistivity, the power $n$ and the critical current density $j_{\mathrm{c}}$ are constant, and $e_0 = 10^{-4}\,\mathrm{Vm}^{-1}$. A different nonlinear current-voltage relation, $\boldsymbol{e}(\boldsymbol{j})$ or $\boldsymbol{e}(\boldsymbol{j},\boldsymbol{h})$, where $\boldsymbol{h}$ is the local magnetic field, can also be employed.

By the Biot-Savart law

$$\boldsymbol{h} = \boldsymbol{h}_{\mathrm{e}} + \boldsymbol{\Phi}[\boldsymbol{j}], \tag{2}$$

where $\boldsymbol{\Phi}[\boldsymbol{j}] = \nabla \times \int_{\Omega} G(\boldsymbol{r} - \boldsymbol{r}')\boldsymbol{j}(\boldsymbol{r}',t)\mathrm{d}\boldsymbol{r}'$, $G(\boldsymbol{r}) = (4\pi|\boldsymbol{r}|)^{-1}$ is the Green function, and $\boldsymbol{r} = (x,y,z)$. Clearly, by the Ampere law, $\nabla \times \boldsymbol{\Phi}[\boldsymbol{j}] = \nabla \times [\boldsymbol{h} - \boldsymbol{h}_{\mathrm{e}}(t)] = \nabla \times \boldsymbol{h} = \boldsymbol{j}$.

To formulate an evolutionary problem for $\boldsymbol{h}$ we use the Faraday law,

$$\mu_0 \dot{\boldsymbol{h}} = -\nabla \times \boldsymbol{e}, \tag{3}$$

where $\mu_0$ is the magnetic permeability of vacuum and the point above a variable means the time derivative. Let at time $t$ the magnetic field $\boldsymbol{h}$ be known. Then

$$\boldsymbol{j} = \nabla \times \boldsymbol{h} \tag{4}$$

and electric field in $\Omega$ is determined by (1). However, $\boldsymbol{e}$ remains undetermined in the non-conductive domain $\Omega_{\mathrm{out}}$. A possible way to deal with this complication is to replace the non-conducting surrounding by a poor conductor obeying, e.g., the linear Ohm law $\boldsymbol{e} = \rho_{\mathrm{out}}\boldsymbol{j}$ with a high resistivity $\rho_{\mathrm{out}}$. Such a solution is not ideal: if $\rho_{\mathrm{out}}$ is high, the evolutionary problem is stiff, otherwise a non-negligible current appears in $\Omega_{\mathrm{out}}$. In [6], the Ohm law is used to define, for each $t$, the electric field in $\Omega_{\mathrm{out}}$ only on the initial step of an iterative determination of $\dot{\boldsymbol{h}}$. After this step the values of $\dot{\boldsymbol{h}}$ in $\Omega$ remain unchanged while in $\Omega_{\mathrm{out}}$ they are modified in the course of iterations ensuring that $\dot{\boldsymbol{j}}_{\mathrm{out}} = \nabla \times \dot{\boldsymbol{h}}|_{\Omega_{\mathrm{out}}} = \boldsymbol{0}$ and thus no stray current in this domain appears. The

role of the fictitious resistivity $\rho_{\text{out}}$ is to efficiently suppress the stray current only in a thin boundary layer outside the superconductor; the iterations take care of this current in the rest of $\Omega_{\text{out}}$.

We now describe the iterative algorithm in detail. To find $\dot{\boldsymbol{h}}$ at time $t$ we compute $\boldsymbol{j}=\nabla\times\boldsymbol{h}$, set

$$\boldsymbol{e}=\begin{cases}\rho(\boldsymbol{j})\boldsymbol{j} & \text{in } \Omega,\\ \rho_{\text{out}}\boldsymbol{j} & \text{in } \Omega_{\text{out}},\end{cases}$$

find $\dot{\boldsymbol{h}}_{\text{in}}=-\mu_0^{-1}\nabla\times\boldsymbol{e}|_{\Omega}$, and define an initial approximation, $\dot{\boldsymbol{h}}_{\text{out}}^{0}$, in $\Omega_{\text{out}}$. Then, on the $i$-th iteration, we compute $\dot{\boldsymbol{j}}^{i}=\nabla\times\dot{\boldsymbol{h}}^{i}$, and set

$$\dot{\boldsymbol{h}}_{\text{out}}^{i+1}=\dot{\boldsymbol{h}}_{\text{e}}+\boldsymbol{\Phi}[\dot{\boldsymbol{j}}_{\text{in}}^{i}]|_{\Omega_{\text{out}}}, \tag{5}$$

where $\dot{\boldsymbol{j}}_{\text{in}}^{i}=\dot{\boldsymbol{j}}^{i}$ in $\Omega$ and zero in $\Omega_{\text{out}}$. Provided these iterations converge, $\nabla\times\dot{\boldsymbol{h}}|_{\Omega_{\text{out}}}=(\dot{\boldsymbol{j}}_{\text{in}})|_{\Omega_{\text{out}}}=\boldsymbol{0}$ as desired. The operator $\boldsymbol{\Phi}$ can be expressed by means of convolutions in $R^3$,

$$\boldsymbol{\Phi}[\boldsymbol{j}]=\begin{Bmatrix} j_z*\partial_y G-j_y*\partial_z G\\ j_x*\partial_z G-j_z*\partial_x G\\ j_y*\partial_x G-j_x*\partial_y G\end{Bmatrix}. \tag{6}$$

Since the Fourier transform of $G$ in $R^3$ is $F(G)=\int_{R^3}G(\boldsymbol{r})e^{-\mathrm{i}\boldsymbol{k}\cdot\boldsymbol{r}}\mathrm{d}\boldsymbol{r}=1/|\boldsymbol{k}|^2$, where $\boldsymbol{k}=(k_x,k_y,k_z)$, see [20, 21], Eq. (6) can be rewritten as

$$\boldsymbol{\Phi}[\boldsymbol{j}]=F^{-1}\left\{\frac{\mathrm{i}}{|\boldsymbol{k}|^2}\begin{bmatrix}k_yF(j_z)-k_zF(j_y)\\ k_zF(j_x)-k_xF(j_z)\\ k_xF(j_y)-k_yF(j_x)\end{bmatrix}\right\}. \tag{7}$$

This expression is not determined for $\boldsymbol{k}=\boldsymbol{0}$ but, taking into account that $\int_{R^3}\boldsymbol{\Phi}[\boldsymbol{j}]\mathrm{d}\boldsymbol{r}=\int_{R^3}[\boldsymbol{h}-\boldsymbol{h}_{\text{e}}(t)]\mathrm{d}\boldsymbol{r}$ should be zero at each moment in time, for $\boldsymbol{k}=\boldsymbol{0}$ we replace $1/|\boldsymbol{k}|^2$ in (7) by zero.

To make this algorithm practical a uniform $N_x\times N_y\times N_z$ grid should be defined in the computational domain $D=\{(x,y,z)|\ \ |x|\le L_x,|y|\le L_y,|z|\le L_z\}$ containing $\Omega$ and some empty space around it; the Fourier transform and its inverse should be replaced by their discrete counterparts on this grid and the FFT algorithm employed. As in [6], we computed the spatial derivatives in (3) and (4) in the Fourier space and applied the Gaussian smoothing; the smoothing parameter $\sigma$ was equal to a half of the grid cell diagonal. The spatially discretized algorithm provides an approximation to $\dot{\boldsymbol{h}}$ values in the grid nodes for a given set of $\boldsymbol{h}$ node values; the resulting system of ordinary differential equations (ODE) was integrated in time by the standard Matlab ODE solver ode23 (with the relative tolerance $2\cdot10^{-4}$). At each time step the iterations (5) were performed until the average of the node values of $|\dot{\boldsymbol{h}}^{i+1}-\dot{\boldsymbol{h}}^{i}|$ in $\Omega_{\text{out}}$ becomes less than $\delta_{\text{it}}\max(|\dot{\boldsymbol{h}}_{\text{e}}|,1)$, where $\delta_{\text{it}}=2\cdot10^{-5}$ was the prescribed tolerance. As the initial approximation $\dot{\boldsymbol{h}}_{\text{out}}^{0}$ we used $\dot{\boldsymbol{h}}_{\text{out}}$ from the previous time step and, usually, the convergence was achieved in a few iterations.

## 3. Simulation results

In our simulations the virgin initial state was assumed and the power $n=30$. We used dimensionless variables,

$$(x', y', z') = \frac{(x, y, z)}{l}, \quad t' = \frac{t}{t_0}, \quad \boldsymbol{e}' = \frac{\boldsymbol{e}}{e_0}, \quad \boldsymbol{j}' = \frac{\boldsymbol{j}}{j_c}, \quad \boldsymbol{h}' = \frac{\boldsymbol{h}}{j_c l}, \tag{8}$$

where $l$ is the characteristic size and $t_0 = \mu_0 j_c l^2 / e_0$; below, the prime is omitted.

Our first example is a hollow, closed at one end superconducting cylinder (Fig 1); its sizes are taken from [2] and scaled in accordance with (8). The cylinder was centrally positioned in a cube with sides 3.2 (the computation domain); in this example we set $\rho_{out} = 20$. We considered two cases. Case A: the external field first grows along the cylinder axis, then in a direction perpendicular to it: $\boldsymbol{h}_e = (0, 0, t)$ for $0 \le t \le \tau$ and $\boldsymbol{h}_e = (t - \tau, 0, \tau)$ for $\tau \le t \le 2\tau$ with $\tau = 0.09$. Case B: same variations in the inverse order, $\boldsymbol{h}_e = (t, 0, 0)$ for $0 \le t \le \tau$ and $\boldsymbol{h}_e = (\tau, 0, t - \tau)$ for $\tau \le t \le 2\tau$. To illustrate the shielding performance, we compared the applied field magnitude $|\boldsymbol{h}_e|$ with the mean magnitude $\langle |\boldsymbol{h}| \rangle$ of the magnetic field inside the "shielded zone" defined as a disk, co-axial to the cylinder, having the thickness 0.143, radius 0.667, and the middle plane at 0.452 below the cylinder top (Fig. 1). Because the cylinder is short, fields perpendicular to its axis penetrate the shielded zone much easier than a field parallel to this axis, see Fig. 2. For the $128 \times 128 \times 128$ grid our computations took about two hours and for the $192 \times 192 \times 192$ grid - twelve hours; the solutions obtained for these grids were close. We present also the calculated distributions of the magnetic field and current density for the case A at $t = \tau$ and $t = 2\tau$ (Fig. 3).

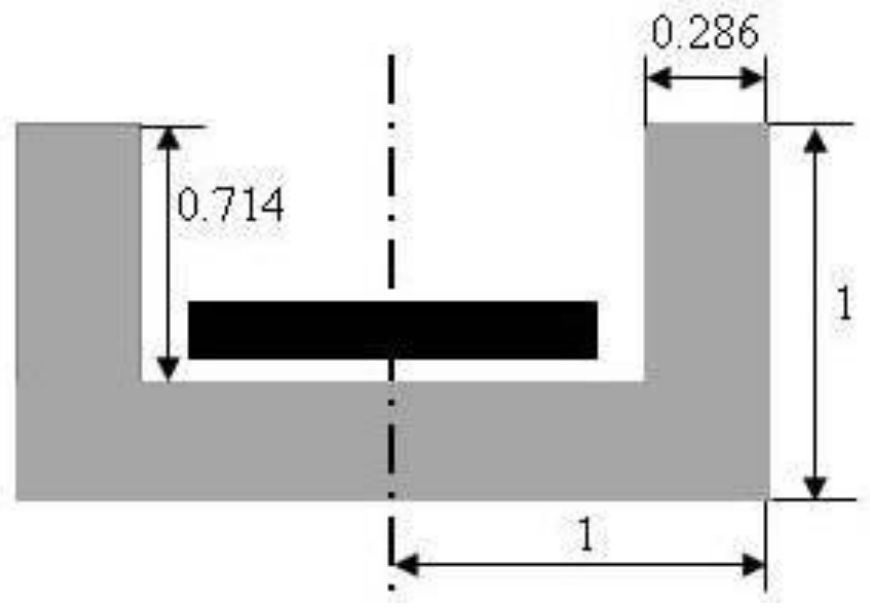

**Figure 1.** Scheme of the superconducting shield (gray); dimensionless units. Black area is the shielded zone.

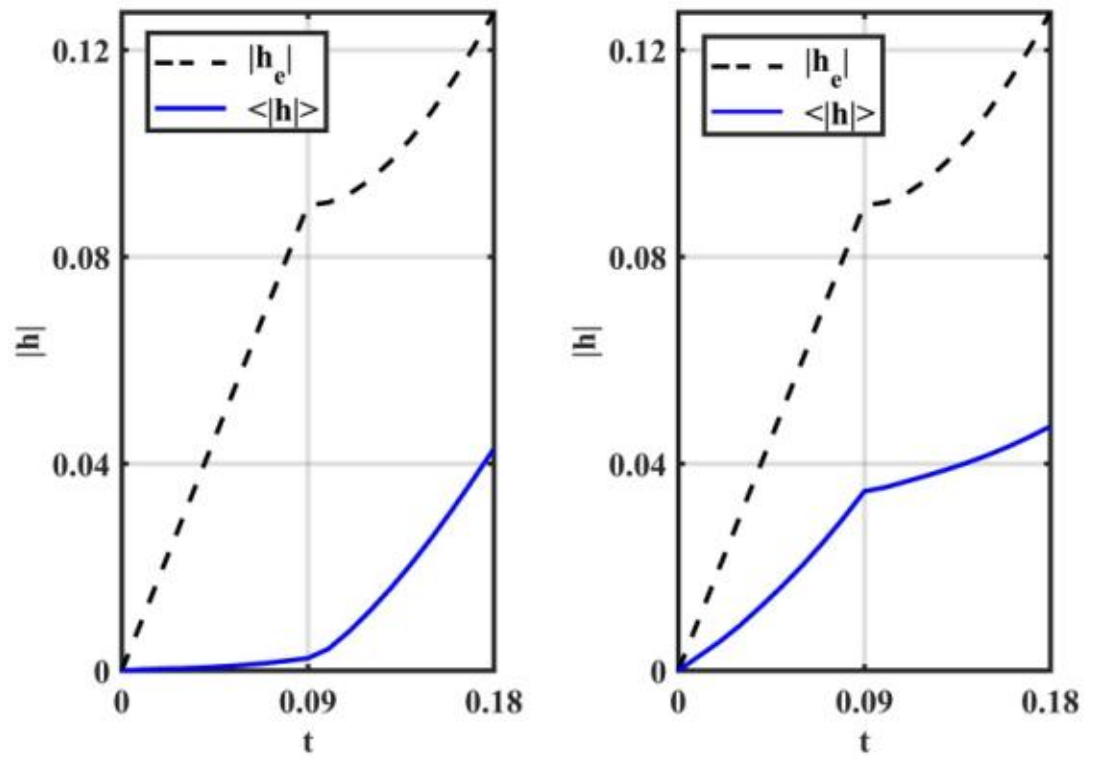

**Figure 2.** Shielding performance, different variations of the external field: left - case A, right - case B. Dashed line - the applied field magnitude, solid line - the average magnetic field magnitude in the shielded zone.

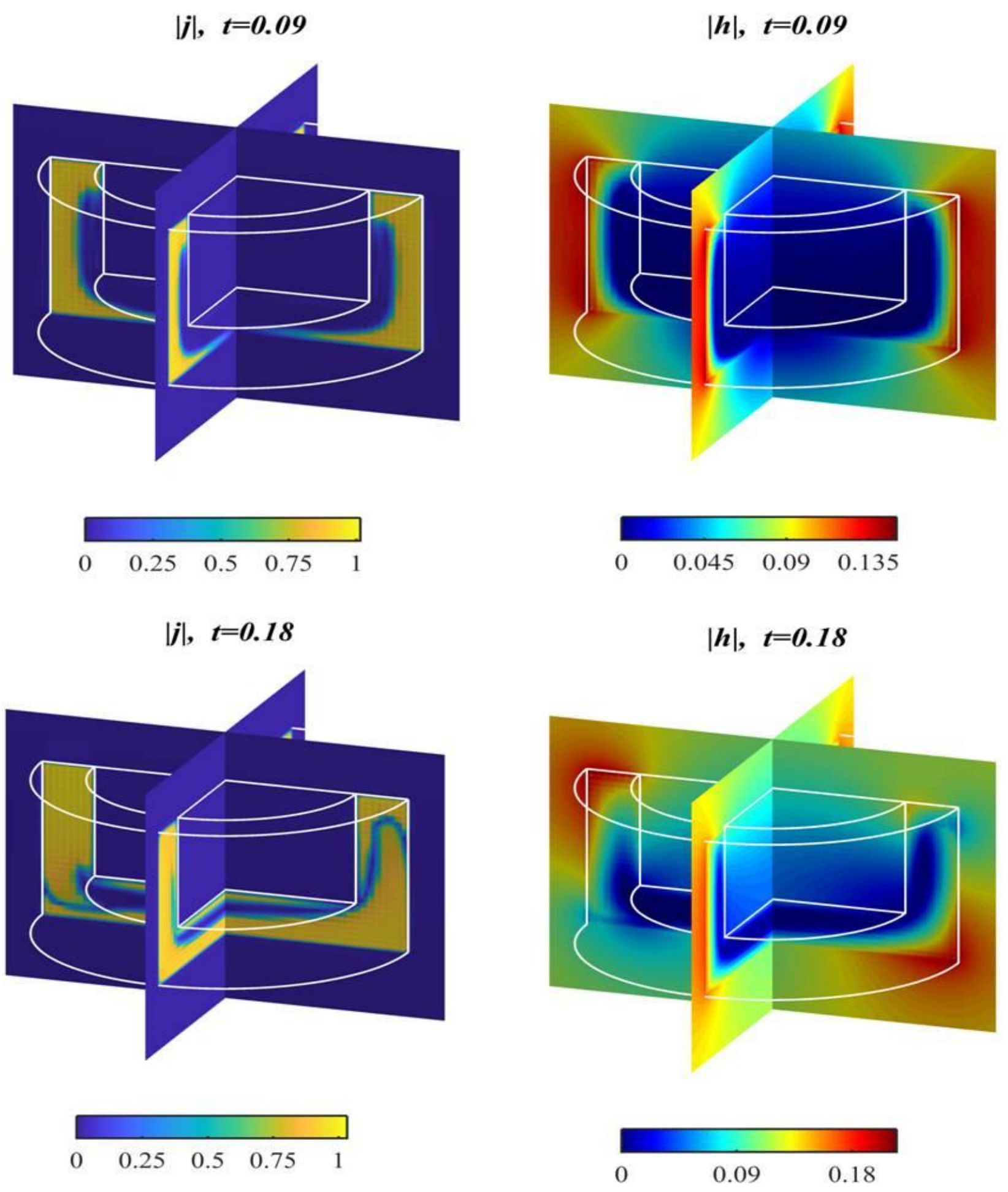

**Figure 3.** Magnetic shield simulation (case A). Magnitudes of the current density (left) and magnetic field (right) in the cross-sections $x=0$ and $y=0$.

The magnetic lens in our second example (Fig. 4) is geometrically similar to those studied experimentally in [13]: it is a superconducting cylinder with a coaxial biconical hole and a narrow slit to prevent the shielding circumference current. The applied field was parallel to the cylinder axis, $\boldsymbol{h}_{\mathrm{e}}=(0,0,t)$, $0\le t\le 0.2$. In this example we used a higher value of the fictitious resistivity, $\rho_{\mathrm{out}}=100$, to fully suppress the flow of current through the narrow slit and the $128\times128\times256$ grid in the computational domain $\{|x|\le1.8,\ \ |y|\le1.8,\ \ |z|\le3.6\}$. The computation time was about six hours. Partially expelled from the superconductor, magnetic field was concentrated in the central part of the lens. The average field magnitude in the central region $\{r\le0.2,\ |z|\le0.33\}$ was about twice bigger than the applied field for $|\boldsymbol{h}_{\mathrm{e}}|=0.05$ and 1.6 times bigger for $|\boldsymbol{h}_{\mathrm{e}}|=0.2$ (Fig. 5). In dimensional units for, e.g., a lens with the outer diameter 30 mm and the critical current density $5\times10^{8}$ A/m$^2$, the field of 0.473 T is amplified twice and of 1.89 T – 1.6 times. The computed current density and magnetic field distributions are presented in Fig. 6 for $t=0.1$ and $t=0.2$.

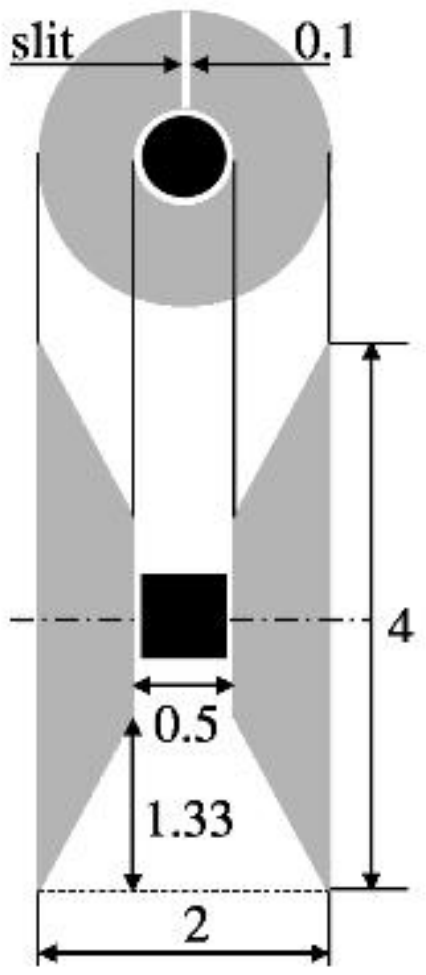

**Figure 4.** Scheme of the magnetic superconducting lens (gray) and the lens central region (black).

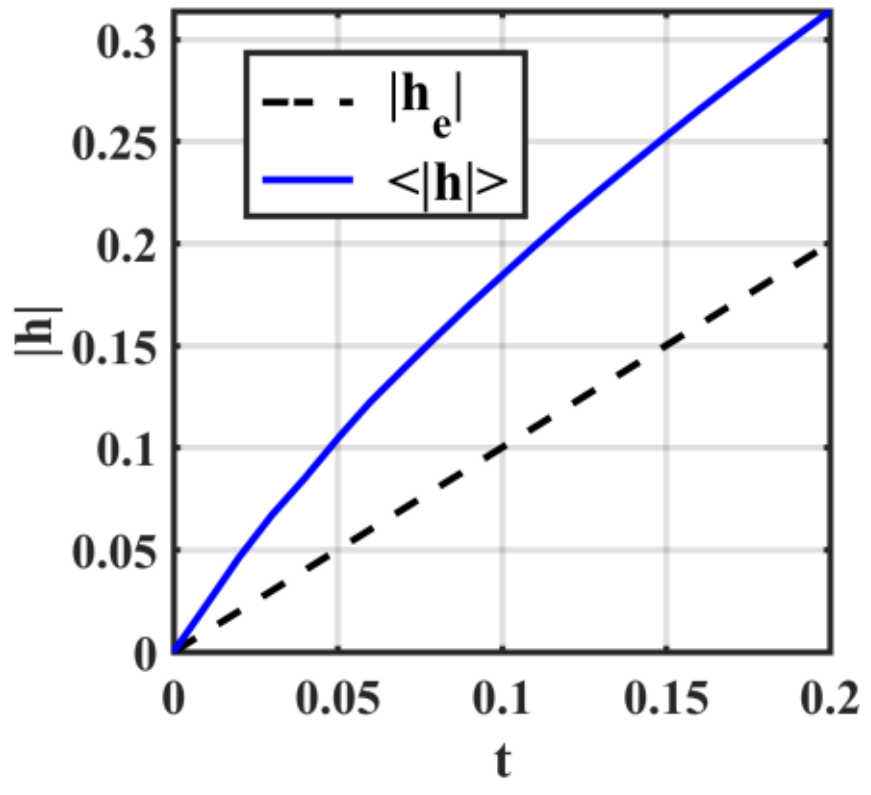

**Figure 5.** Magnetic lens. Dashed line - the applied field magnitude, solid line - the average magnetic field magnitude in the lens central region.

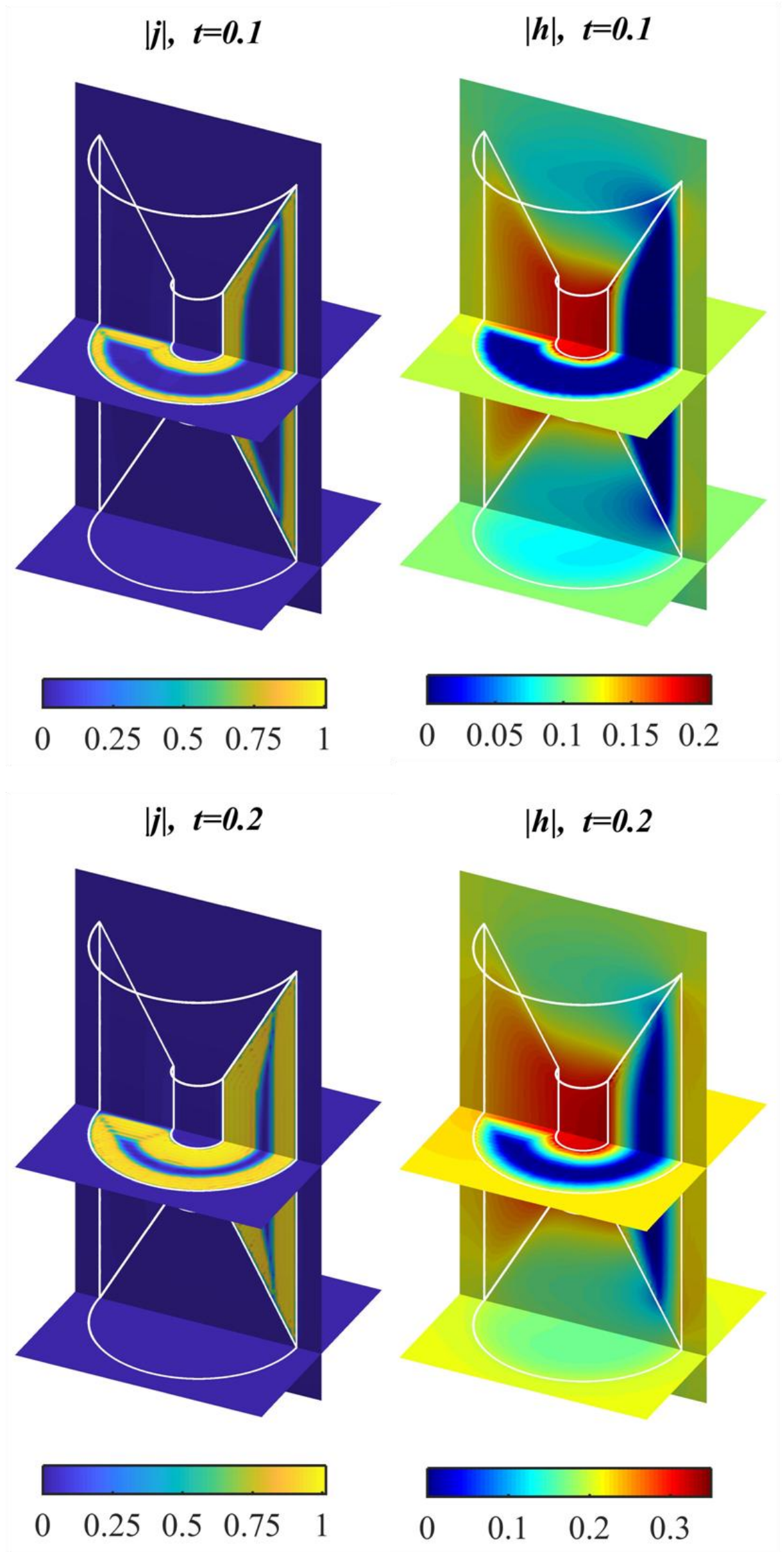

**Figure 6.** Superconducting magnetic lens. Magnitudes of the current density (left) and magnetic field (right) in the cross-sections $x=0$, $z=0$, and $z=-2$.

## Conclusion

In many cases design of superconductor devices needs solution of highly nonlinear 3D electromagnetic problems. Development of numerical methods for these problems is currently an active field of research.
Presented solutions of two problems, shielding and concentration of magnetic field by bulk superconductors, confirm the efficiency of the 3D FFT-based method [6] and illustrate its ability to solve realistic magnetization problems for superconductors of different shapes. The method is simple, easy to implement, and can be used with any current-voltage relation characterizing superconducting material (iso- or anisotropic, field dependent, etc.) Jointly with the homogenized anisotropic bulk model, this method can be used also for modeling magnetization of superconducting film stacks, a promising alternative to the bulk superconductor magnets.